\documentclass{article}

\usepackage{PRIMEarxiv}

\usepackage[utf8]{inputenc} % allow utf-8 input
\usepackage[T1]{fontenc}    % use 8-bit T1 fonts
\usepackage{url}            % simple URL typesetting
\usepackage{booktabs}       % professional-quality tables
\usepackage{amsfonts}       % blackboard math symbols
\usepackage{nicefrac}       % compact symbols for 1/2, etc.
\usepackage{microtype}      % microtypography
\usepackage{lipsum}
\usepackage{fancyhdr}       % header
\usepackage{graphicx}       % graphics
\graphicspath{{media/}}     % organize your images and other figures under media/ folder
\usepackage{times}
\usepackage{graphicx}
\usepackage[colorlinks=true,linkcolor=blue,citecolor=blue,allcolors=blue]{hyperref}
\usepackage{amssymb} % For checkmark and xmark
\usepackage{pifont} % For checkmark and xmark
\usepackage{tabularx} % For checkmark and xmark
\usepackage{graphicx} % For checkmark and xmark
\usepackage{todonotes}
\usepackage{comment}
\usepackage{multirow}
\usepackage{listings}
\usepackage{footnote}
\usepackage{subcaption}
\makesavenoteenv{tabular}
\newcommand{\cmark}{\ding{51}}% For checkmark and xmark
\newcommand{\xmark}{}% For checkmark and xmark
\title{Scholarly Wikidata: Population and Exploration of Conference Data in Wikidata using LLMs}

\vspace{-2cm}

\author{
  Nandana Mihindukulasooriya \\
  IBM Research\\
  New York, USA\\
  \texttt{nandana@ibm.com} \\
   \And
  Sanju Tiwari \\
  Sharda University \\
  Greater Noida, India \\
  \texttt{tiwarisanju18@ieee.org} \\
  \And
  Daniil Dobriy \\
  Vienna University for Economics and Business \\
  Vienna, Austria \\
  \texttt{daniil.dobriy@wu.ac.at} \\  
  \And
  Finn Årup Nielsen \\
DTU Compute, Technical University of Denmark \\
Lyngby, Denmark\\
  \texttt{faan@dtu.dk}  
  \And
  Tek Raj Chhetri \\
McGovern Institute for Brain Research, Massachusetts Institute of Technology \\ | Center for Artificial Intelligence (AI) Research Nepal \\
Cambridge, USA | Sundarharaincha-09, Nepal\\
  \texttt{tekraj.chhetri@cair-nepal.org}  
  \And
  Axel Polleres \\
  Vienna University for Economics and Business \\
  Vienna, Austria \\
  \texttt{axel.polleres@wu.ac.at} \\   
}

\begin{document}
\maketitle

\vspace{-1cm}

\begin{abstract}
% Scholarly data, including information about conferences, scientific papers, and authors, serves as a vital resource for diverse stakeholders, such as students, researchers, conference organizers, sponsors, and funding agencies, %among others. 
 Several initiatives have been undertaken to conceptually model the domain of scholarly data using ontologies and to create respective Knowledge Graphs.
 %of scholarly data. 
 Yet, the full potential seems unleashed, as automated means for automatic population of said ontologies are lacking, and respective initiatives from the Semantic Web community are not necessarily connected: we propose to make scholarly data more sustainably accessible by leveraging Wikidata's infrastructure and %tool ecosystem and 
 automating its population in a sustainable manner through LLMs by tapping into unstructured sources like conference Web sites and proceedings texts as well as already existing structured conference datasets. While an initial analysis shows that Semantic Web conferences are only minimally represented in Wikidata, we argue that our methodology can help to populate, evolve and maintain scholarly data as a community within Wikidata. 
 % Thus, our work focuses on populating Wikidata with scholarly data
 % extracting information from structured and unstructured data sources.

 Our main contributions include (a) an analysis of ontologies for representing scholarly data to identify gaps and relevant entities/properties in Wikidata, (b) semi-automated extraction -- requiring (minimal) manual validation -- of conference metadata (e.g., acceptance rates, organizer roles, programme committee members, best paper awards, keynotes, and sponsors) from websites and proceedings texts using LLMs. Finally, we discuss (c) extensions to visualization tools in the Wikidata context for data exploration of the generated scholarly data. Our study focuses on data from 105 Semantic Web-related conferences and extends/adds more than 6000 entities in Wikidata. It is important to note that the method can be more generally applicable beyond Semantic Web-related conferences for enhancing Wikidata's utility as a comprehensive scholarly resource.
%\textbf{Resource Type:} Knowledge Graph\\
\\
~\textbf{Source Repository: } \url{https://github.com/scholarly-wikidata/} \\
~\textbf{DOI:} \url{https://doi.org/10.5281/zenodo.10989709} \\
~\textbf{License:  Creative Commons CC0 (Data), MIT (Code)} 
\end{abstract}


\section{Introduction}\label{sec:intro}
Scientific conferences are vital for researchers to share their research findings and advancements. It offers an opportunity to discuss research problems or limitations, a platform for networking with peers, and a platform for promoting collaboration, which is essential for learning, innovation, and problem-solving. Because of the importance of scientific conferences, we have seen tremendous growth in the number of conferences over the years~\cite{10.1007/978-3-030-30796-7_6}. For example, IEEE (Institute of Electrical and Electronics Engineers) sponsors more than 2,000\footnote{\url{https://www.ieee.org/about/at-a-glance.html}} conferences and events annually. Similarly, ACM (Association for Computing Machinery) hosts more than 170\footnote{\url{https://www.acm.org/conferences/about-conferences}} conferences annually worldwide.

\begin{sloppypar}
Therefore, efforts have been made to capture metadata about scientific events~\cite{10.1007/978-3-030-30796-7_8,fathalla2019eventskg,10.1007/978-3-030-30796-7_6,nuzzolese2016conferencea} in a linked-data format as they provide valuable information.
Such data can be used for (i) better understanding the progress of science overall, (ii) the evolution of particular research topics (or fields), (iii) understanding research impact (e.g. by sponsors' interest) over time, etc. The availability of scholarly metadata enables scientometrics~\cite{kirr-etal-2020SWJ_SWdecade}, or practical tools such as recommending relevant conferences or papers to readers~\cite{10.1007/978-3-030-30796-7_8} for navigating through the fastly growing scientific output which is becoming time-consuming and almost impractical.
\end{sloppypar}
 
However, as much as the benefits these metadata about scientific events provide, there exist challenges. The primary obstacle is the collection of large-scale metadata, which is nontrivial in nature~\cite{10.1007/978-3-030-30796-7_8}. Similarly, the sustainability, which is also the focus of this paper, of the accumulated metadata constitutes the second and most significant obstacle. If the data collected is not sustainable, it may be lost over time, resulting in the loss of valuable information and efforts put into data collection. For instance, the Microsoft Academic Graph, which contained over 8 billion triples~\cite{10.1007/978-3-030-30796-7_8} with information about scientific publications and related data, was retired in December 2021\footnote{\url{https://www.microsoft.com/en-us/research/project/microsoft-academic-graph/}}. While the effort was somewhat continued shortly later in OpenAlex\footnote{\url{https://openalex.org/}}, the case demonstrates sustainability issues in individual or commercial scholarly KG offerings. 

We argue that collaborative, general purpuse, community-driven platforms, such as Wikipedia, are generally more sustainable than such fragmented efforts: community participation is motivated by intrinsic factors, fostering a sense of belonging to the group~\cite{XU2015275}. Notably, commercial initiaves seem to recognize this, as shown by Google's declaration that it will cease operations on Freebase and transfer its content to Wikidata~\cite{10.1145/2872427.2874809}. Wikidata, which focuses on knowledge graphs (KGs), is a sister project of Wikipedia and another example of a community-driven platform~\cite{10.1007/978-3-319-11964-9_4,10.1145/2629489}. Wikidata currently has more 110M entities and 25K active contributors\footnote{\url{https://www.wikidata.org/wiki/Wikidata:Statistics}}. By bringing Scholarly data about scientific conferences into Wikidata, they can be seamlessly integrated with existing background knowledge through SPARQL queries. Furthermore, Wikidata benefits from a robust tooling ecosystem and widely used libraries, including entity linkers, search tools, SPARQL endpoint with high-availability, easy-to-use query editor, visualization tools, and more~\cite{DiefenbachWA21,rossenova2022wikidata}. Wikidata also allows non-expert users to directly access the KGs through search and Web UI (user interface). Therefore, the primary objective of our work is to integrate scientific conference metadata into Wikidata, a community-led platform. 

After conducting an analysis of Wikidata entities related to Semantic Web conferences such as International Semantic Web Conference (ISWC), Extended/European Semantic Web Conference (ESWC), International Conference Knowledge Engineering and Knowledge Management (EKAW), International Conference on Knowledge Capture (K-CAP), SEMANTiCS, and Knowledge Graph and Semantic Web Conference (KGSWC), it was noticed that some conferences were missing and the ones that were present had only minimal information. In this project, we have extended Wikidata to include a more comprehensive set of information (e.g. see ISWC 2023\footnote{\url{https://www.wikidata.org/wiki/Q119153957}} (Q119153957)). Within the scope of this work, we focused on the Semantic Web conferences but our method is more generally applicable and can be extended to other conference series. We note that 105 conferences we added to, updated in Wikidata is higher than the comparable related work such as Scholarly Data (35 confs)\footnote{\url{https://bit.ly/3Vs6XNc}}, ORKG (5 confs)~\footnote{\url{https://orkg.org/organizations/Event}} as of July, 2024.

Large language models (LLMs) have proven their language understanding capabilities with many NLP benchmarks~\cite{min2023recent}. In recent years, approaches such as in-context learning with a few-shot example have allowed them to perform many tasks such as relation or fact extraction~\cite{Khorashadizadeh23,MihindukulasooriyaTEL23}. Such models can be used to easily extract information from sources with natural language text, such as conference proceedings, websites, or call for papers. Nevertheless, their output can be prone to errors. In our work, LLMs are used to extract data, which is then verified by a human-in-the-loop validation to eliminate any noisy extraction and ensure accuracy.

In particular, this paper makes the following contributions.
\begin{itemize}
    \item We analysed existing ontologies for representing scholarly data and mapped them to Wikidata to identify relevant Wikidata entities/properties as well as gaps. 
    \item We present a method for utilizing large language models to efficiently extract conference metadata from various sources, curating them through a human-in-the-loop validation process using OpenRefine, and populating the data in Wikidata via Wikidata QuickStatements and provide an evaluation for LLM-based extractions. 
    \item As a result of this project, we have extended over 1000 existing entities and created more than 5,000 new entities, including conferences, scientific articles, and people. These entities are now available on Wikidata and can be accessed via the Web UI or SPARQL endpoint.
    \item We extend visualization tools  Scholia\footnote{\url{https://scholia.toolforge.org/}} and Synia\footnote{\url{https://synia.toolforge.org}} to better visualize the information we added to Wikidata. 
\end{itemize}

\section{Related Work}
\label{sec:related}

Scientific events have emerged as a crucial element in scholarly communication across various scientific fields. They serve as central hubs for fostering scientific connections among various elements such as individuals (e.g., organizers and attendees), locations, activities (e.g., participant roles), and materials (e.g., conference proceedings) within the realm of scholarly discourse \cite{10.1007/978-3-030-30796-7_6}. 

This section will explore the existing work in the related topic.

Different works have been done to capture and (re-)use the metadata about the scholarly events.  The first work is by Fathalla et al.~\cite{10.1007/978-3-030-30796-7_6} who developed Scientific Events Ontology (OR-SEO) to capture the information of scientific events. OR-SEO is currently utilized as the framework for event pages on the OpenResearch\footnote{\url{http://openresearch.org.}} (OR) platform, which serves as a semantic wiki platform aimed at crowdsourcing metadata and supporting scholarly metadata management and utilization. Similarly, Fathalla et al.~\cite{fathalla2019eventskg}, in their other work, introduce a 5-star Linked Dataset containing dereferenceable IRIs (Internationalized Resource Identifier), which is an updated version of the EVENTSKG dataset. The dataset contains information about 73 different conferences in the field of computer science, such as Artificial Intelligence (AI),  World Wide Web (WEB), and Software Engineering (SE). Nuzzolese et al.~\cite{nuzzolese2016semantic} examine the Semantic Web Dog Food (SWDF) dataset and explore its quality and sustainability challenges. The SWDF employs the Semantic Web Conference (SWC) ontology as the foundational ontology for representing data related to academic conferences. Proposed approach uses cLODg3 (conference Linked Open Data generator)~\cite{nuzzolese2016generating} to regenerate the SWDF dataset based on the conference ontology, offering a sustainable solution. Scholarly data~\cite{nuzzolese2016conferencea} initiative refactors SWDF to continue the growth of the dataset. It introduces the conference ontology, which improves SWC. The Open Research Knowledge Graph (ORKG) is one of the main knowledge graphs designed to enable the structured representation and sharing of scholarly knowledge. ORKG aims to transform scholarly publications into a structured, interconnected knowledge graph, allowing for easier data access, analysis, and reuse. Currently, ORKG has 5 conference records covering 3 conference series under the event.class\footnote{\url{https://orkg.org/organizations/Event}}.

In addition to scholarly ontologies, work has been done on scholarly information tooling, such as visualization and scraping. Angioni et al.~\cite{Angioni2022AIDA}, for example, developed an AIDA dashboard for analyzing and comparing scientific conferences. Their work uses Computer Science Ontology and the AIDA knowledge graph which enable the construction of visualization, such as top authors and organizations. Similarly, Kruger et al.~\cite{Kruger2000DEADLINER} developed an early system named DEADLINER to extract information, such as deadline, topic, title, and program committee, from conference and workshop announcements (call for papers). The other similar work is by Fahl et al.~\cite{Fahl2023Semantification} who implemented a system for scraping the CEUR Workshop Proceedings site. Fahl et al.'s~\cite{Fahl2023Semantification} work also includes entity linking of event locations and proceeding editor disambiguation. Kirrane et. al. \cite{kirr-etal-2020SWJ_SWdecade} has presented a qualitative analysis of the main seminal papers by adopting a top-down approach and produced results with three bottom-up data-driven approaches (Saffron, Rexplore, PoolParty). The analysis has been conducted on the corpus of Semantic Web papers acquired from 2006 to 2015. Several other efforts already done in earlier time and proposed Bibster, a Peer-to-Peer system \cite{haase2004bibster} for transforming the bibliographic data among research community. 
Scholia, which our work re-uses and extends, is another application by Nielsen et al.~\cite{NielsenF2017Scholia} that allows visualization of scholarly profiles for topics, people, organizations, species, chemicals, etc., using bibliographic and other information in Wikidata. In addition to visualization, Scholia also has functionality for scraping data from the DOI (Digital Object Identifier), NeurIPS conferences, CEUR workshop series, and Open Journal Systems. Synia is a Scholia-inspired system that creates profiles based on Wikidata, but with templates defined on a wiki. Some of Synia's profiles are displaying scholarly information \cite{Nielsen2023Synia}. 

Moreover, there have been several other related initiatives. Wikicite\footnote{\url{https://meta.wikimedia.org/wiki/WikiCite}} is one such initiative, which focuses on a bibliographic database of source and citation metadata~\cite{Taraborelli2016WikiCite}. This work benefits from some outcomes from Wikicite such as ``\emph{is proceedings from}'' (P4745) property. Efforts have also been made to maintain conference acceptance rates, such as Open Research\footnote{\url{https://www.openresearch.org/wiki/ISWC}} or Conference-Acceptance-Rate\footnote{\url{https://github.com/lixin4ever/Conference-Acceptance-Rate}}, which this work makes re-use of.

\section{Overview of Scholarly Wikidata Process}
The objective of our work is to extract information from structured and unstructured data sources and add it to Wikidata, specifically related to Semantic Web conferences. We leverage existing structured data and tools thereby promoting the reusability principles of the Semantic Web and further enhance them with additional information extracted from text using LLMs.

\begin{figure}[htbp]
  \centering
  \includegraphics[width=1\textwidth]{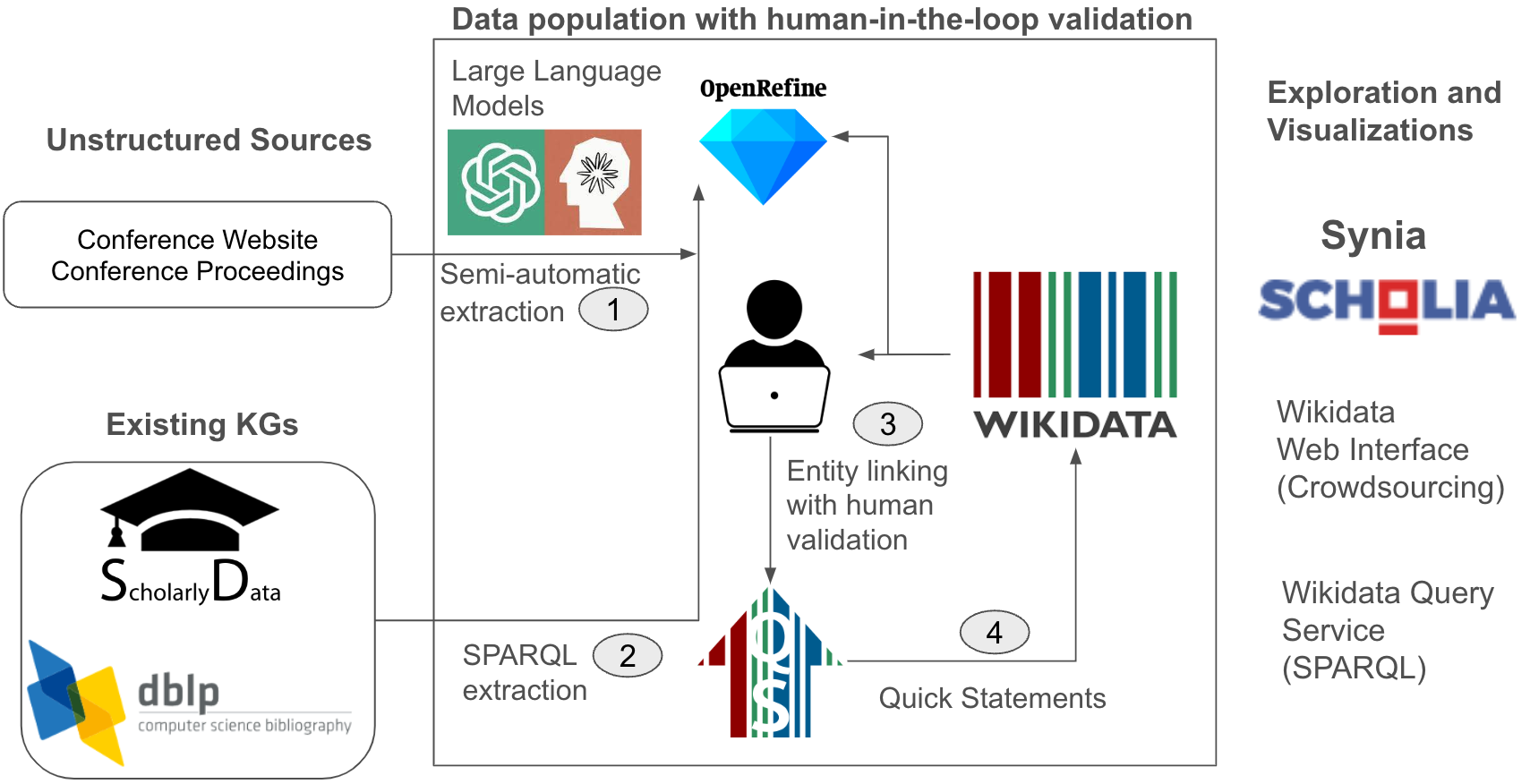}
  \caption{Overview of the methodology. The code used for the process including the LLM prompts, SPARQL queries, OpenRefine schemas are available in the GitHub repository.}
  \label{fig:workflow}
\end{figure}

\subsection{Data Sources}
We considered both structured and unstructured data as input to our process. We wanted to extract information from authoritative sources, so we used the official conference proceedings and the conference website as input to our process. Table~\ref{tab:data_sources} illustrates the different sources we used in our process. These sources contained most of the information we wanted to extract.  If the data has already been extracted and exposed in structured formats, especially in RDF (Resource Description Framework), such as DBLP and Scholarly Data, we reuse them. 

\begin{table}[th]
\centering
\caption{Information extracted from different sources}
\label{tab:data_sources}
\begin{tabular}{|l|p{13cm}|}
\hline
\textbf{Data source} & \textbf{Extracted Information}                                                                  \\ \hline
\begin{tabular}[c]{@{}l@{}}conference \\ proceedings \\ front matter\end{tabular} &
  \begin{tabular}[c]{@{}l@{}}organization committee and their roles, number of papers submitted, accepted for each track \\ of the conference, programme committees for each track with the track and their roles \\ (PC, SPC), and prominent topics of submitted papers\end{tabular} \\ \hline
conference website   & important dates for each track (deadlines), number of attendee                                  \\ \hline
DBLP KG              & papers published in the main and poster/demo tracks along with their authors and other metadata \\ \hline
Scholarly Data KG    & events such as tutorials, workshops, panels, and keynotes                                       \\ \hline
\end{tabular}
\end{table}

\subsection{Data population with human-in-the-loop validation}

Figure \ref{fig:workflow} shows the process we followed to extract the data from different sources and populate data in Wikidata. 

\paragraph{1. semi-automatic extraction from unstructured sources}
In order to extract the necessary information from unstructured sources such as conference proceedings front matter or conference websites, we have LLMs using the LangChain\footnote{\url{https://www.langchain.com/langchain}} framework. For conference front matter, the PDFs are loaded with the ``PyPDFLoader'' and converted into semantically meaningful chunks first using the section headings, and then using ``SemanticChunker''\footnote{\url{https://python.langchain.com/docs/modules/data_connection/document_transformers/semantic-chunker/}}. For conference websites, given the URL, the crawler we implemented navigates to the website and traverses the contents. The main functions of the crawler are: (a) retrieving a sitemap if it exists and extending it through traversing the links, (b) extracting embedded structured data (Microdata, JSON-LD, RDFa, OpenGraph), and (c) retrieving HTML and text contents of the website pages. In both cases, relevant chunks for each extraction task are selected based on the section heading and a set of pre-defined keywords. 

Finally, we formulated LLM prompts for specific extraction tasks such as extraction of the number of submitted/accepted papers for calculating the acceptance rate for each track, organization committee with their roles, programme committee with track, and roles, and conference deadlines. The Appendix\footnote{link provided at the end of the paper} illustrates an example for extracting submitted/accepted papers from the proceedings front matter using in-context learning capabilities with two-shot examples. The LLM outputs are parsed to extract the output in CSV format which is used as input to OpenRefine. For further details, please refer to the implementation in GitHub repository.  

\paragraph{2. SPARQL query extraction from structured sources}
DBLP provides their data as an RDF dump with monthly releases~\footnote{\url{https://blog.dblp.org/2022/03/02/dblp-in-rdf/}}, and we loaded the April 2024 snapshot into a local Virtuoso instance to create a SPARQL endpoint. Scholarly Data makes an SPARQL endpoint\footnote{\url{http://www.scholarlydata.org/sparql/}} available online. After analyzing the ontology used and the data available in each of those KGs, we formulated SPARQL queries to extract papers for each conference including both main tracks and posters and demos from DBLP and sub-events of conferences such as workshops, tutorials, panels, and keynotes from Scholarly Data. An example SPARQL query is shown in Appendix.

\paragraph{3. Entity linking with human validation}
In order to avoid creating duplicate entities and properly link to existing entities, we used the entity reconciliation functionality of OpenRefine. This step also involves disambiguation of authors having the same name. DBLP optionally provides Wikidata ID, ORCID and Google Scholar IDs when available which helps to disambiguate the authors better. Before populating Wikidata, data accuracy is ensured through human-in-the-loop validation performed using Open Refine UI.

\paragraph{4. Wikidata population using Quick Statements}
We defined mappings between the CSV data and Wikidata using Open Refine schema definitions. The Appendix shows an example of such a definition. Finally, Open Refine allowed us to create Quick Statements that can be imported as a batch to Wikidata. Please refer to \cite{mihin2024} for details.

\subsection{Exploration and visualizations}

\begin{table}[th]
\caption{Examples of Synia templates and examples of their corresponding pages}
\label{tab:synia}
\begin{tabular}{|p{2cm}|p{6.5cm}|p{6.5cm}|}
\hline
Description                         & Synia Template & Synia Page \\ \hline
Scientific event              & \url{https://www.wikidata.org/wiki/Wikidata:Synia:scientificevent}       &   \url{https://synia.toolforge.org/#scientificevent/Q119153957}         \\ \hline
Scientific events index       & \url{https://www.wikidata.org/wiki/Wikidata:Synia:scientificevent-index} & \url{https://synia.toolforge.org/#scientificevent/}\\ \hline
Scientific Event series       & \url{https://www.wikidata.org/wiki/Wikidata:Synia:scientificeventseries} & \url{https://synia.toolforge.org/\#scientificevent/Q119153957} \\ \hline
Scientific event series index & \url{https://www.wikidata.org/wiki/Wikidata:Synia:scientificeventseries-index} & \url{https://synia.toolforge.org/\#scientificeventseries} \\ \hline
\end{tabular}
\end{table}

We have used Scholia and Synia as visualization tools to allow the community to explore conference-related information with visual summaries, including line charts (e.g, acceptance rates, number of participants), area charts (e.g, topics through time), maps (e.g, conference locations), graphs (e.g., co-author graphs), and timelines (e.g., important dates, deadlines).  While Scholia's visualizations of Wikidata content are defined in the Scholia Web application with Jinja2 templates, Synia's visualizations are defined on wiki pages, making it easier to add new visualizations. The visualizations are generated reusing the standard graph plotting of the Wikidata Query Service and embedded on the Scholia (or Synia) webpage via Iframes. To visualize the new conference data we created new Synia template wiki pages. Table~\ref{tab:synia} illustrates some Synia templates and their corresponding pages. Some SPARQL queries in the Synia templates were modified from the equivalent ones in Scholia.

\begin{sloppypar}
We also added a template with faceted views on organizational roles. For example, 
\href{https://synia.toolforge.org/\#scientificeventseries/Q6053150/organizationalrole/Q125207931}{\texttt{https://synia.toolforge.org/\#scientificeventseries/-}} \\ \href{https://synia.toolforge.org/\#scientificeventseries/Q6053150/organizationalrole/Q125207931}{\texttt{Q6053150/organizationalrole/Q125207931}} displays information about the Semantic Web challenge chairs through the history of ISWC. By changing the role, say for example, research track chair (Q125207937) or sponsor chair (Q125207972), similar pages can be generated for any organizer role. 
\end{sloppypar}

\begin{sloppypar}
Some of the template definitions of Synia were transfered to Scholia, e.g., 
\url{https://scholia.toolforge.org/event/Q119153957} now displays number of parti\-cipants and acceptance rate for the ISWC 2023 conference.
\end{sloppypar}

\section{Mapping Conference Ontologies to Wikidata}

Several ontologies representing conference metadata exist in literature as discussed in Section~\ref{sec:related}. Our objective was to use existing ontologies to understand the important aspects that should be represented in Wikidata. We utilized those conceptual models to identify the relevant Wikidata entities, properties, and qualifiers to represent scholarly data. We identified gaps in Wikidata and proceeded to add the missing elements to the platform to extend its capabilities.

\begin{table}[h]
\centering
\caption{Ontologies and their coverage of conference aspects}
\label{tab:ontologies}
\begin{tabular}{l|ccccccccccccc|}
\textbf{Ontology} & \rotatebox{80}{\textbf{Conf. Metadata}} & \rotatebox{80}{\textbf{Conf. Series}} & \rotatebox{80}{\textbf{Topical Coverage}} & \rotatebox{80}{\textbf{Roles, Committees}} & \rotatebox{80}{\textbf{Sub-events}} & \rotatebox{80}{\textbf{Publications}} & \rotatebox{80}{\textbf{Awards}} & \rotatebox{80}{\textbf{Registration}} & \rotatebox{80}{\textbf{Tracks}} & \rotatebox{80}{\textbf{Sponsors}} & \rotatebox{80}{\textbf{Submission, Review}}  & \rotatebox{80}{\textbf{Attendance}} \\
\hline
Wikidata & \cmark & \cmark & \cmark & \cmark & \cmark & \cmark & \cmark & \xmark & \cmark & \cmark & \cmark  & \cmark \\
\hline
Scholarly Data & \cmark & \cmark & \cmark & \cmark & \cmark & \cmark & \cmark & \xmark & \cmark & \cmark & \cmark  & \xmark \\
\hline
SEO (ontology) & \cmark & \xmark & \xmark & \cmark & \cmark & \xmark & \cmark & \cmark & \xmark & \cmark & \cmark  & \xmark \\
\hline
SemWeb Conference & \cmark & \xmark & \xmark & \cmark & \cmark & \xmark & \xmark & \xmark & \xmark & \xmark & \xmark &  \xmark \\
\hline
SciGraph & \cmark & \cmark & \cmark & \xmark & \cmark & \cmark & \xmark & \xmark & \cmark & \xmark & \xmark  & \xmark \\
\hline
Schema.org & \cmark & \xmark & \cmark & \cmark & \cmark & \cmark & \cmark & \xmark & \xmark & \cmark & \cmark& \xmark \\
% \hline
\end{tabular}
\end{table}

\subsection{Overview of Mappings to Wikidata}

This section provides an overview of how existing conference ontology properties maps to Wikidata properties. Table \ref{tab:ontologies} shows a comparison summary of several conference-related ontologies and Wikidata. More detailed mappings are documented in the Wiki\footnote{\url{https://github.com/scholarly-wikidata/scholarly-wikidata/wiki/Mapping-scholarly-data-ontologies-to-Wikidata}} of the project.

Based on this analysis, we identified several properties that are missing in Wikidata. For example, our proposals on ``number of submissions''\footnote{\url{https://www.wikidata.org/wiki/Wikidata:Property_proposal/number_of_submissions}} and ``number of accepted contribution''\footnote{\url{https://www.wikidata.org/wiki/Wikidata:Property_proposal/number_of_accepted_contributions}} were accepted by Wikidata.

\subsection{Extensions to qualifier values}

\begin{figure}[tbp]
  \centering
  \includegraphics[width=1\textwidth]{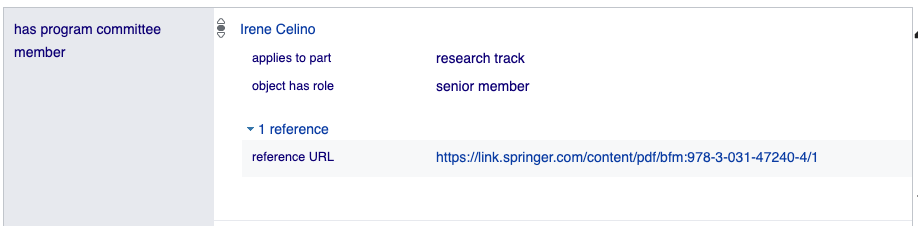}
  \caption{An example use of qualifiers in ISWC 2023 Wikidata entity (Q119153957).}
  \label{fig:qualifier-example}
\end{figure}

Wikidata data model has a flexible way to include fine-grained information through a primary relation and a set of qualifiers. Fig.~\ref{fig:qualifier-example} illustrated how the property ``has programme committee member'' property is used to link ``Irene Celino'' to ``ISWC 2023'' and qualifiers ``applies to'' and  ``object has role'' are used to provide additional information that she was an SPC in the research track. We used Reference URL to refer to the source where the fact was extracted. Based on our analysis of the existing ontologies in literature as discussed before, we have added several qualifier values to populate richer information. Table~\ref{tab:qual} illustrates some examples of such additions. For instance, more than 60 organizer roles were added to represent the conferences we analyzed. In addition to adding new qualifiers, we also modified the property constraints appropriately considering the use case of conference metadata.

\begin{table}[h!]
\caption{Qualifer values (entities) added to Wikidata based on analysis of existing scholarly data ontologies to represent the conference metadata.}
\label{tab:qual}
\begin{tabular}{|p{4cm}|p{1.2cm}|p{2cm}|p{7.5cm}|}
\hline
property &
  qualifer &
  parent entity &
  entities (only labels are shown for brevity) \\ \hline
significant event (P793) &
   point in time (P585) & 
  deadline (Q2404808) &
  abstract submission deadline, paper submission deadline, acceptance notification deadline,  camera-ready submission deadline \\ \hline
organizer (P664) &
  object has role (P3831) &
  role (Q4897819) &
  programme chair, organization chair, workshop chair, local chair and 60 others\footnote{\url{https://w.wiki/9mWB}} \\ \hline
sponsor (P859) &
  object has role (P3831) &
  sponsorship level (Q117280318) &
  diamond sponsor, platinum sponsor, gold plus sponsor, gold sponsor, silver plus sponsor, silver sponsor \\ \hline
winner (P1346) &
  object has role (P3831) &
  best paper award (Q112270830) &
  best research track paper award,  best research track student paper award, best resource track paper award, best demo paper award, best poster paper award \\ \hline
has program committee member (P5804), admission rate (P5822) &
  applies to part (P518) &
  conference track (Q66087801) &
  research track, resources track, in-use track,  posters and demos track, position paper track, evaluations and experiments track\\ \hline
\end{tabular}
\end{table}

\section{Data exploration}

\begin{figure}[btp]
  \centering
  \includegraphics[width=1\textwidth]{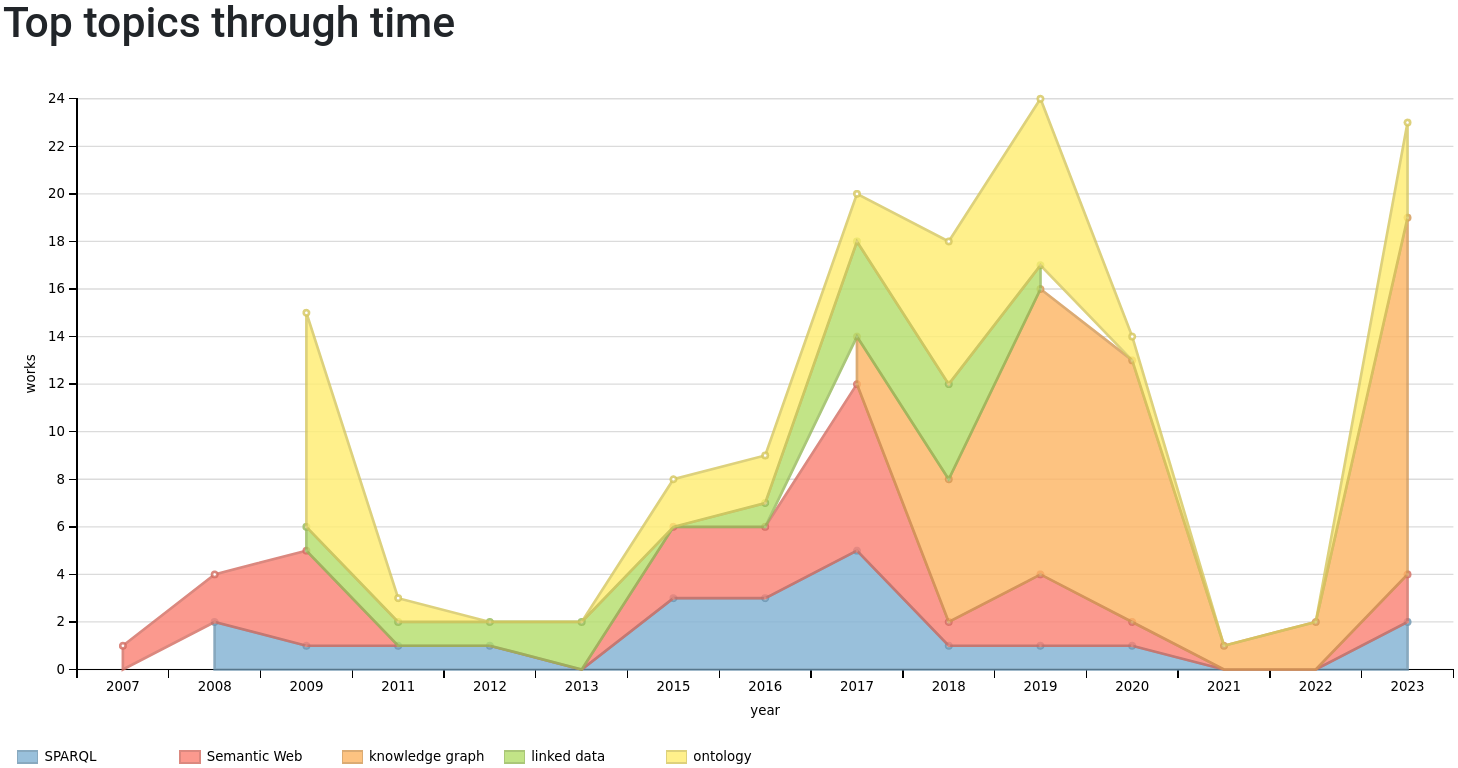}
  \caption{Screenshot from Scholia for a panel on the page for the ESWC conference series \url{https://scholia.toolforge.org/event-series/Q17012957\#top-topic-through-time}.}
  \label{fig:eswc-top-topics-through-time}
\end{figure}

\begin{figure}[tbp]
  \centering
  \includegraphics[width=1\textwidth]{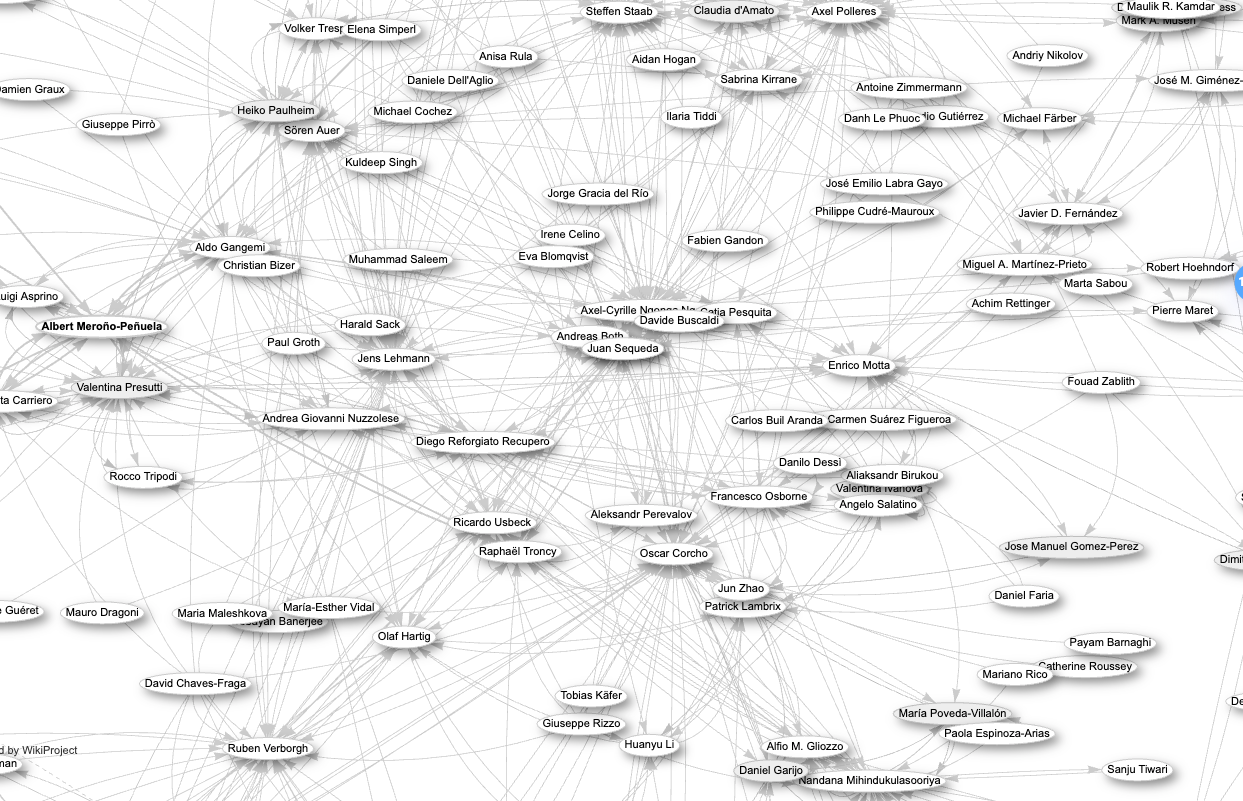}
  \caption{Screenshot from Scholia for the co-author graph panel on the page for the ISWC 2023 conference. All co-authorships are shown, also those beyond the specific conference. It shows only part of the graph and the photographs of people have been removed (due to copyright). For the interactive complete graph, please refer to \url{https://scholia.toolforge.org/event/Q119153957\#co-authors}.}
  \label{fig:iswc-2023-coauthor-graph}
\end{figure}

Fig.~\ref{fig:eswc-top-topics-through-time} shows a screenshot of Scholia displaying the topics though time for a scientific event series, the ESWC conference series.
The SPARQL query behind this visualization assembles topics for each year based on the chain event series - event - proceedings - article - topic and the publication year of the article.
This visualization was inspired by a note in \cite{Angioni2022Leveraging}:
``One drawback of [Scholia] is that the topics are associated to venues as a whole and cannot be used to evaluate their temporal evolution''. Any interpretation of the plots should be aware that the data behind the plot is not complete, i.e., not all conference articles are added and annotated in Wikidata.
In Fig.~\ref{fig:iswc-2023-coauthor-graph} another Scholia screenshot shows the co-author graph of people associated with the ISWC conference series.

\newpage
\begin{figure}[!htbp]
  \centering
  \begin{subfigure}[b]{1\textwidth}
    \includegraphics[width=1\textwidth]{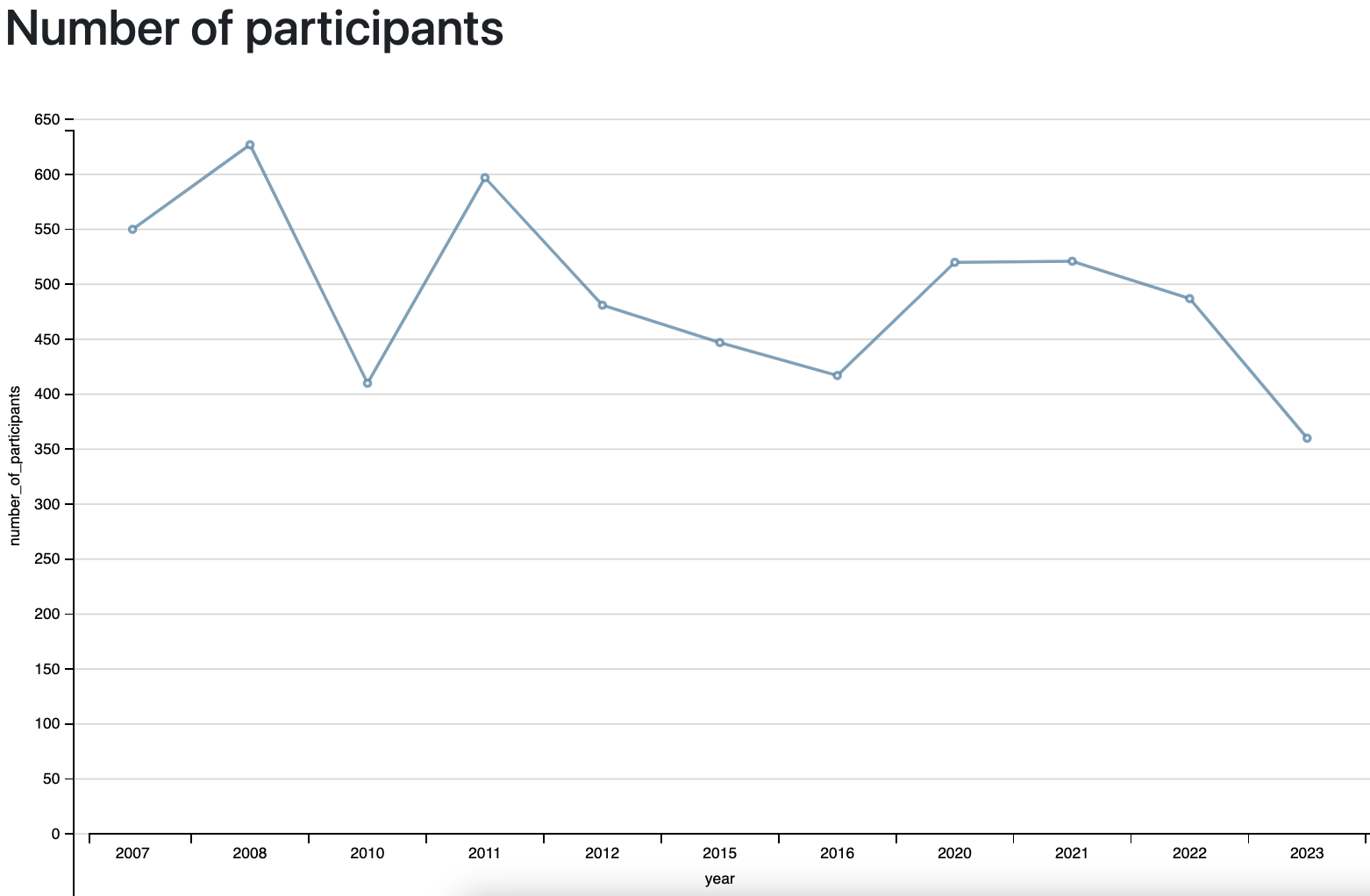}
    \label{fig:neurips-number-of-participants}
  \end{subfigure}
  \begin{subfigure}[b]{1\textwidth}
    \includegraphics[width=1\textwidth]{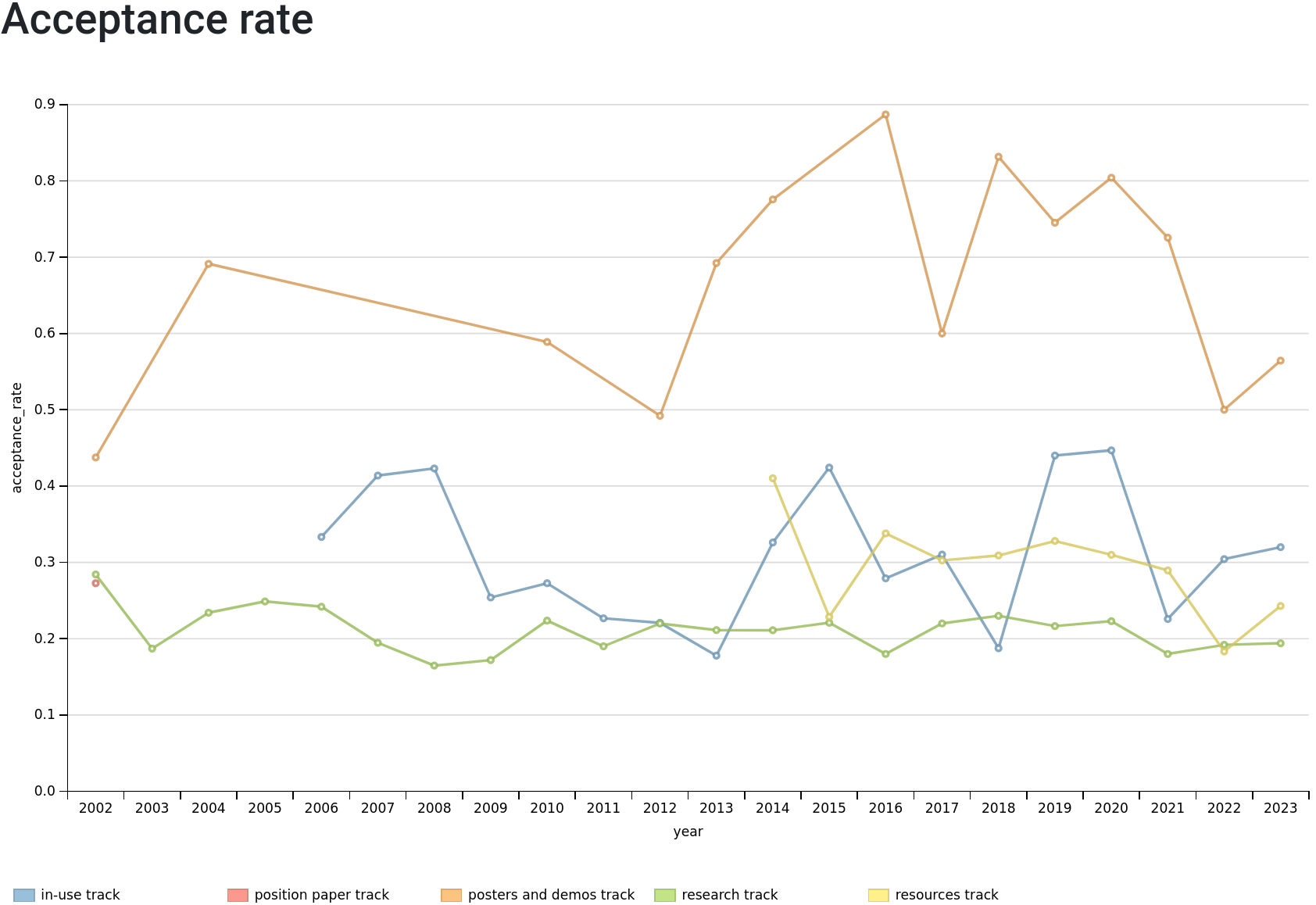}
    \label{fig:iswc-acceptance-rate}
  \end{subfigure}
  \caption{Screenshots from Synia showing the number of participants and acceptance rates of ISWC through time. From \url{https://synia.toolforge.org/\#scientificeventseries/Q6053150}.}
  \label{fig:charts}
  
  \vspace{2mm} % Adjust this value to control space between subfigures and the next image
  
  \includegraphics[width=1\textwidth]{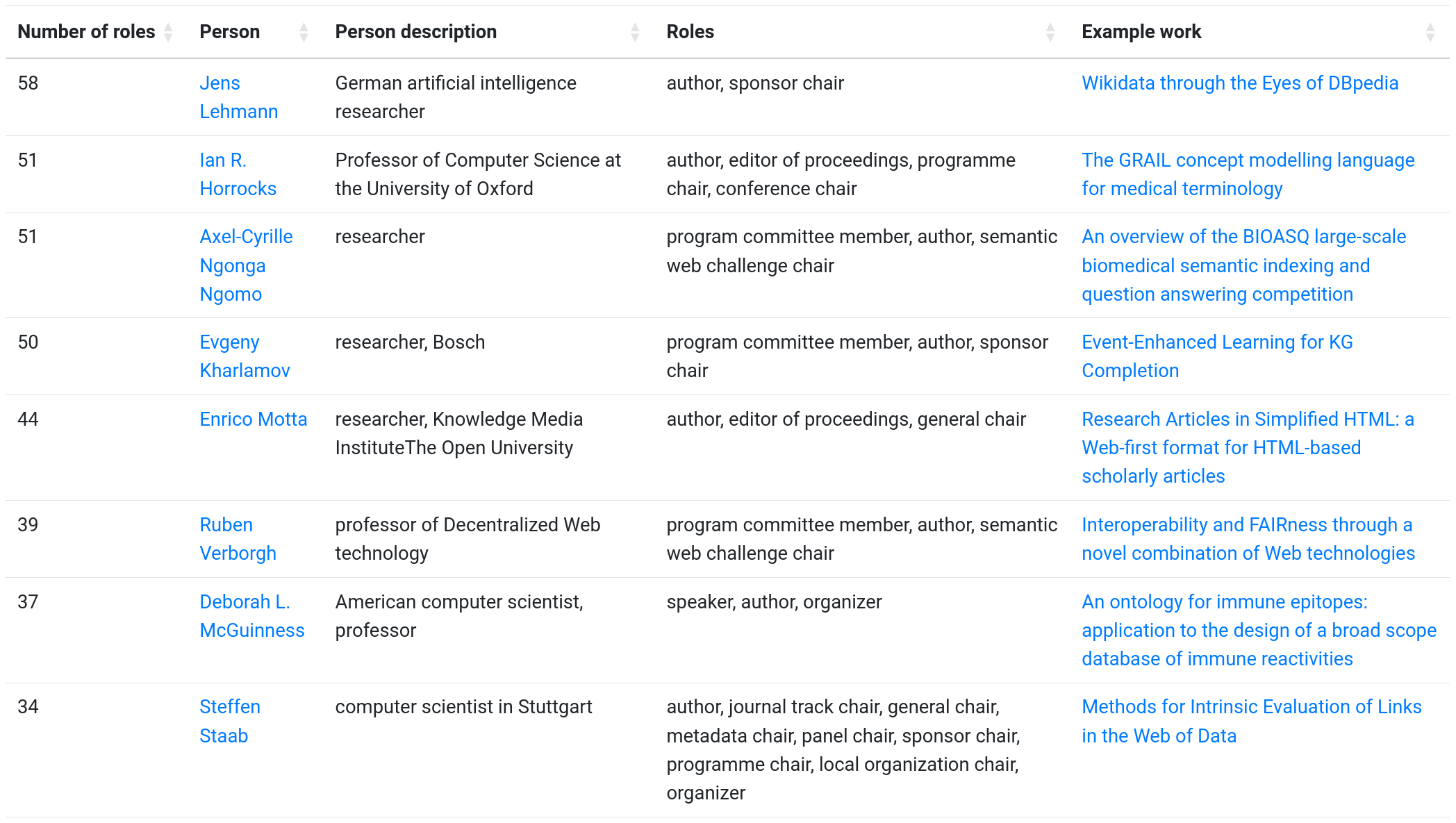}
  \caption{Screenshot from Synia of table with people associated with ISWC from \url{https://synia.toolforge.org/\#scientificeventseries/Q6053150}.}
  \label{fig:iswc-people}
\end{figure}
\par\noindent

% \vspace{-2mm} 

For visualizing the new conference data added, we created new Synia templates in Wikidata for scientific events and scientific event series with tables and visualizations panels.
Some of Synia panels were also incorporated into Scholia.
Among the panels we added were line charts for the number of participants and the acceptance rate through time for a conference series. Fig~\ref{fig:charts} illustrates the participants and acceptance rate through time for ISWC.

Fig.~\ref{fig:iswc-people} is a screenshot from a part of the Synia page about the ISWC conference series listing eight top people associated with the conference (e.g., as author, organizer or speaker) and ordered according to number of roles  with unique roles displayed in the fourth column. This column shows the information from the ``object has role'' qualifiers for the organizer value, --- if available in Wikidata.
This listing shows an unequal distribution of roles where some researchers have been in several different organizational roles throughout the years while others have mostly participated in the conference as authors.

\section{Evaluation}

In this section, we present evaluations for four extraction tasks using two LLMs to extract facts from preface proceedings and conference websites.
\paragraph{Setup} 
For this evaluation, we used text from 42 prefaces and websites (22 ISWC, 20 ESWC) as a test dataset. For each task, we created a human extraction ground truth by manually extracting the facts by reading the preface. Then after performing the LLM extractions, we compared system generations with the human ground truth to calculate micro precision, recall, and F1 metrics on extraction. The system is not penalized for not extracting the information that is not available in the text.

\paragraph{Discussion} 
Table~\ref{tab:eval} shows that both LLMs we tested performed well in extracting submitted/accepted papers, organizational roles, and PC members from the proceeding front matter.  The proceedings clearly list the organization roles and PC members and LLMs can extract them with high accuracy. On rare occasions, LLM confuses organizations (affiliations) with people. Extraction of the number of papers submitted/accepted in each track requires more natural language understanding. It was noted that sometimes the LLM extracted correct facts from its pre-trained knowledge in addition to the provided text. For example, for ISWC 2007, it generated that 7 tutorials were accepted with 1 invited tutorial even though the preface never mentioned 7 tutorials. Another example, from ``\emph{The main scientiﬁc program of ESWC 2020 contained 39 papers: 26 papers in the research track, 8 papers in the resources track, and 5 papers in the in-use track. The papers were selected out of 166 paper submissions, with a total acceptance rate of 23.5\% (22\% for the research track, 26\% for the resources track, and 28\% for the in-use track).}``, claude-3 extracted (research: 26/119, resources: 8/31, in-use: 5/18) which coincide with the acceptance ratios but with a total of 168 submissions. GPT-4 only extracted the number of accepted papers. 

\begin{table}[h]
\centering
\caption{Evaluation Results}
\label{tab:eval}
\begin{tabular}{|l|l|l|llc|lcc|}
\hline
\multirow{2}{*}{Task} &
  \multirow{2}{*}{Source} &
  \multirow{2}{*}{Model} &
  \multicolumn{3}{l|}{ISWC (2002-2023)} &
  \multicolumn{3}{l|}{ESWC (2004-2023)} \\ \cline{4-9} 
 &
   &
   &
  \multicolumn{1}{l|}{Precision} &
  \multicolumn{1}{l|}{Recall} &
  \multicolumn{1}{l|}{F1} &
  \multicolumn{1}{l|}{Precision} &
  \multicolumn{1}{l|}{Recall} &
  \multicolumn{1}{l|}{F1} \\ \hline
\multirow{2}{*}{\begin{tabular}[c]{@{}l@{}}Submitted/accepted\\ papers extraction\end{tabular}} &
  \multirow{2}{*}{\begin{tabular}[c]{@{}l@{}}Proceedings\\ Front Matter\end{tabular}} &
  gpt-4 &
  \multicolumn{1}{c|}{0.92} &
  \multicolumn{1}{c|}{0.87} &
  0.89 &
  \multicolumn{1}{c|}{0.94} &
  \multicolumn{1}{c|}{0.95} &
  0.95 \\ \cline{3-9} 
 &
   &
  claude-3 &
  \multicolumn{1}{c|}{1.00} &
  \multicolumn{1}{c|}{0.89} &
  0.93 &
  \multicolumn{1}{c|}{0.90} &
  \multicolumn{1}{c|}{0.97} &
   0.93\\ \hline
\multirow{2}{*}{\begin{tabular}[c]{@{}l@{}}Organization role \\ extraction\end{tabular}} &
  \multirow{2}{*}{\begin{tabular}[c]{@{}l@{}}Proceedings\\ Front Matter\end{tabular}} &
  gpt-4 &
  \multicolumn{1}{c|}{1.00} &
  \multicolumn{1}{c|}{1.00} &
  1.00 &
  \multicolumn{1}{c|}{1.00} &
  \multicolumn{1}{c|}{1.00} &
  1.00 \\ \cline{3-9} 
 &
   &
  claude-3 &
  \multicolumn{1}{c|}{1.00} &
  \multicolumn{1}{c|}{0.96} &
  0.98 &
  \multicolumn{1}{c|}{1.00} &
  \multicolumn{1}{c|}{0.92} &
  0.96 \\ \hline
\multirow{2}{*}{\begin{tabular}[c]{@{}l@{}}PC member \\ extraction\end{tabular}} &
  \multirow{2}{*}{\begin{tabular}[c]{@{}l@{}}Proceedings\\ Front Matter\end{tabular}} &
  gpt-4 &
  \multicolumn{1}{c|}{1.00} &
  \multicolumn{1}{c|}{1.00} &
  1.00 &
  \multicolumn{1}{c|}{1.00} &
  \multicolumn{1}{c|}{1.00} &
  1.00 \\ \cline{3-9} 
 &
   &
  claude-3 &
  \multicolumn{1}{c|}{0.98} &
  \multicolumn{1}{c|}{1.00} &
  \multicolumn{1}{c|}{0.99} &
  \multicolumn{1}{c|}{0.99} &
  \multicolumn{1}{c|}{1.00} &
   0.99\\ \hline
\multirow{2}{*}{\begin{tabular}[c]{@{}l@{}}Important dates\\ extraction\end{tabular}} &
  \multirow{2}{*}{Website} &
  gpt-4 &
  \multicolumn{1}{c|}{0.8} &
  \multicolumn{1}{c|}{0.97} &
  \multicolumn{1}{c|}{0.88} &
  \multicolumn{1}{c|}{0.81} &
  \multicolumn{1}{c|}{0.94} &
  \multicolumn{1}{c|}{0.87} \\ \cline{3-9} 
 &
   &
  claude-3 &
  \multicolumn{1}{c|}{0.28} &
  \multicolumn{1}{c|}{0.92} &
  \multicolumn{1}{c|}{0.43} &
  \multicolumn{1}{c|}{0.07} &
  \multicolumn{1}{c|}{0.81} &
  \multicolumn{1}{c|}{0.13} \\ \hline
\end{tabular}
\end{table}

When extracting deadlines from websites, we noticed some hallucinations with the Claude-3 model, which extracts dates and deadlines that are not mentioned on the web page or wrongly categorizing the deadlines (e.g., identifying deadlines for poster CFPs as deadlines for full paper CFPs), resulting in a lower precision value of 28\% for ISWC websites and even 7\% for EWSC websites. Still, even gpt-4 could only achieve the precision of 80\% and 81\%, for ISWC and ESWC conferences respectively. The recall values for both models range from 81\% to 97\%, with claude-3 achieving the lowest performance with regard to ESWC websites (81\%), mostly owing to the fact that most deadlines are identified, even though it is miscategorized. As future work, we will work on strategies to mitigate these issues with improved prompt engineering or other techniques such as chain-of-thoughts or self-verification~\cite{WengZX0HLSLZ23}.

\section{Conclusions and Future Work}
In this paper, we present our work on scholarly data, with a focus on Semantic Web-related conferences. The paper addresses the sustainability challenges (see Section~\ref{sec:intro}) of metadata by bringing the metadata of scholarly events, particularly conferences, to Wikidata. Our work uses LLMs to efficiently extract data from authoritative sources such as conference proceedings and websites and performs a human-in-the-loop validation to ensure any noisy extractions are eliminated. Our work has resulted in the creation of more than 6K entities or more than 30k statements about scientific conferences in Wikidata, providing valuable information about Semantic Web-related conferences in a machine-processable format aka KGs representation.

% In conclusion, we have extended the scholarly data in Wikidata, especially focused on Semantic Web-related conferences. We used LLMs to efficiently extract data from authoritative sources such as conference proceedings and websites and performed a human-in-the-loop validation to ensure any noisy extractions were eliminated. More than 6K entities are created or extended with more than 30K statements.

\paragraph{Impact and Reusability}
By populating scholarly data in Wikidata, we make them available to a large community of both experts and non-experts using SPARQL endpoints, Web UI, and other tools such as visualizations or entity linkers. Wikidata is supported by the Wikimedia Foundation and a large community of active contributers, ensuring its sustainability. The availability of this scholarly data in Wikidata enables the community to easily access information such as acceptance rates by making simple queries. Organizers and track chairs can easily access the programme committee and other information from past years. By combining with existing data, interested parties can run ad-hoc queries, for example, the gender balance in ISWC organization committees throughout the previous years (see query\footnote{\url{https://w.wiki/9nnJ}}). The data we added to Wikidata, along with their sources, can potentially be used to generate benchmarks for tasks in scholarly domain, such as information extraction or question answering.

% There is now a gender panel in Synia: https://synia.toolforge.org/#scientificeventseries/Q6053150

The data in Wikidata follows Linked Data and Findable, Accessible, Interoperable, and Reusable (FAIR) principles. 
In addition to the visualizations in Scholia and Synia, we also provide a Wiki page with a list of example queries\footnote{\url{https://github.com/scholarly-wikidata/scholarly-wikidata/wiki/Scholary-Wikidata-Query-Examples}}.

% Does the resource break new ground?
% Does the resource fill an important gap?
% How does the resource advance the state of the art?
% Has the resource been compared to other existing resources (if any) of similar scope?
% Is the resource of interest to the semantic web community?
% Is the resource of interest to society in general?
% Will/has the resource have/had an impact, especially in supporting the adoption of semantic web technologies?

% Is there evidence of usage by a wider community beyond the resource creators or their project? Alternatively (for new resources), what is the resource’s potential for being (re)used?
% Is the resource easy to (re)use? For example, does it have high-quality documentation? Are there tutorials available?
% Is the resource general enough to be applied in a wider set of scenarios, not just for the originally designed use? If it is specific, is there substantial demand?
% Is there potential for extensibility to meet future requirements?
% Does the resource include a clear explanation of how others use the data and software? Or (for new resources) how others are expected to use the data and software?
% Does the resource description clearly state what the resource can and cannot do, and the rationale for the exclusion of some functionality?

\paragraph{Design \& Technical Quality} 
We analyzed the existing ontologies in the scholarly data domain and used them to identify appropriate Wikidata entities, properties, and qualifiers. We followed the best practices of Wikidata by using appropirate qualifiers. For the properties we identified as gaps in Wikidata, for example, the number of submitted/accepted papers, we created property proposals following Wikidata guidelines as discussed in Section 4.1. We used authoritative sources such as conference proceedings or official website as our source. Furthermore, we included the reference URLs of sources where we extracted the facts to ensure their provenance is maintained and the community members can verify those. We used tools such as OpenRefine to perform entity reconciliation to avoid creation of duplicates when some entities already existed in Wikidata. 

%
% - Does the design of the resource follow resource-specific best practices? YES, permanent URI, 
% - Did the authors perform an appropriate reuse or extension of suitable high-quality resources? For example, in the case of ontologies, authors might extend upper ontologies and/or reuse ontology design patterns.
% - Is the resource suitable for solving the task at hand? YES -> argue
% - Does the resource provide an appropriate description (both human- and machine-readable), thus encouraging the adoption of FAIR principles? Is there a schema diagram? For datasets, is the description available in terms of VoID/DCAT/DublinCore?

\paragraph{Availability}
The data we generated is added to Wikidata and is already accessible via the Wikidata web user interface and SPARQL endpoint. Data in Wikidata is distributed under a Creative Commons CC0 license. Additionally, the source code for the extraction and population process is the GitHub repository under MIT license. A snapshot of the proceeding front matter links, website crawls, SPARQL query results, LLM extractions, and evaluation benchmarks are available on Zenodo, with a Creative Commons CC0 license, a persistent URI (DOI) and a canonical citation.

As future work, we are planning on creating a comprehensive benchmark for extracting triples from scholarly data communications with manually curated ground truth and making it available for the community. We believe such benchmark can help the community to evaluate new approaches and strategies for information extraction from sources using novel technologies such as LLMs.

Furthermore, based on our empirical evaluations based on the human-created ground truth, we plan to recommend a subset of extraction tasks that can be fully automated with LLMs with minimal noise so that they can be integrated to Wikidata as bots to perform large-scale extractions on all conference series.

The fact that Wikidata can be edited by anyone is quite useful for crowd-sourcing information, but also brings some challenges. 
Even though our experience with scholarly data on Wikidata since 2016 does not show its a common occurrence, 
Wikidata can be prone to vandalism and inclusion of falsified information by malicious parties~\cite{DBLP:conf/cikm/HeindorfPSE16}. Wikidata tracks the provenance of all edits with username / IP addresses, and analysis of these edits using ML models can be used to detect vandalism. Ensuring the vera\-city (i.e., accuracy) of crowd-sourced conference information within Wikidata poses an open challenge and we plan to formulate different strategies as part of our future work.

%Appendix of the paper is available here.~\footnote{\url{https://github.com/scholarly-wikidata/scholarly-wikidata/blob/fa6bbdc78f69df81ae12b45d8537e1977eee8aa6/docs/EKAW_2024_Paper_Appendix.pdf}}

% Does the design of the resource follow resource-specific best practices?
% Did the authors perform an appropriate reuse or extension of suitable high-quality resources? For example, in the case of ontologies, authors might extend upper ontologies and/or reuse ontology design patterns.
% Is the resource suitable for solving the task at hand?
% Does the resource provide an appropriate description (both human- and machine-readable), thus encouraging the adoption of FAIR principles? Is there a schema diagram? For datasets, is the description available in terms of VoID/DCAT/DublinCore?

\section*{Acknowledgements}
This research was funded in whole or in part by the Austrian Science Fund (FWF) [10.55776/COE12].

%Bibliography
\bibliographystyle{unsrt}  
\bibliography{references}  

\newpage
\appendix

\vbox{%
    \hsize\textwidth
    \linewidth\hsize
    \vskip 0.1in
  \hrule height 4pt
  \vskip 0.25in
  \vskip -5.5pt%
  \centering
    {\LARGE\bf{Appendix} \par}
      \vskip 0.29in
  \vskip -5.5pt
  \hrule height 1pt
  \vskip 0.09in%

  }

\section{An example of prompt used to extract the number of submitted/accepted papers}\label{app:prompts_ex}

It contains two messages for the chat models where the first one is the system message and the second one is the human message. Furthermore, it contains two examples (two-shots) to specify the task as well as the output format. The \{preface\_text\} variable in line 27 is replaced with the conference proceedings text from which the submitted/accepted paper counts to be extracted.

\lstset{
    language=Python,
    basicstyle=\small,
    numbers=left,
    numberstyle=\tiny,
    numbersep=5pt,
    frame=single,
    breaklines=true,
    breakatwhitespace=true,
    tabsize=4,
    showstringspaces=false,
    extendedchars=true,
    literate={ﬁ}{{fi}}1 {ﬂ}{{fl}}1 {ﬀ}{{ff}}1,
}
\begin{lstlisting}

acceptance_ratio_system_template = ``You are a data entry clerk and you know how to read some text and extract the requested information. You always follow the expected output format shown in the examples precisely. You only extract information in the text and do not introduce any new facts."

acceptance_ratio_human_template = ``Given a preface from a conference proceedings, extract the number of research papers submitted and accepted to each track. Output the results in the CSV format.

Here are two examples:

Preface: The main scientiﬁc program of ESWC 2023 contained 41 papers selected out of 167 submissions (98 research, 23 in-use, 46 resource): 19 papers in the research track, 9 in the in-use track, and 13 in the resource track. The overall acceptance rate was 24% (19% research, 39% in-use, 28% resource).

Output:
track, submitted, accepted
research, 98, 19
in-use, 23, 9
resource, 46, 13
--- complete ----

Preface: The research papers program received 245 full paper submissions, which were ﬁrst evaluated by the Program Committees of the respective tracks. The review process included evaluation by Program Committee members, discussions to resolve conﬂicts, and a metareview for each potentially acceptable borderline submission. After this a physical meeting among Track and Conference Chairs was organized to see that comparable evaluation criteria in diﬀerent tracks had been used and to discuss remaining borderline papers.As a result, 52 research papers were selected to be presented at the conference. The proceedings also include ten PhD symposium papers presented at a separate track preceding the main conference, and 17 demo papers giving a brief description of the system demos that were accepted for presentation in a dedicated session during the conference.

Output:
track, submitted, accepted
research, 245, 52
PhD symposium, - , 10
demo, - , 17
--- complete ----

Now extract the number of research papers submitted and accepted to each track from the following text. Output only the CSV content and "--- complete ----" as the last line.

Preface: {preface_text}"

\end{lstlisting}

\section{SPARQL queries used to extract the papers of a given proceedings and their corresponding authors from the DBLP KG.}\label{app:query_ex}
\lstset{
    language=Python,
    basicstyle=\small,
    numbers=left,
    numberstyle=\tiny,
    numbersep=5pt,
    frame=single,
    breaklines=true,
    breakatwhitespace=true,
    tabsize=4,
    showstringspaces=false
}
% \begin{lstlisting}

\begin{figure}[h]
%\footnotesize
\begin{lstlisting}[frame=single,breaklines]{sparql}

# get the list of papers in a given proceeding 
#identifier <__proceedings_uri__>

PREFIX dblp: <https://dblp.org/rdf/schema#> 

SELECT ?paper ?title ?doi ?pages ?year WHERE {
    ?paper a dblp:Publication, dblp:Inproceedings;
        dblp:title ?title;
        dblp:doi ?doi;
        dblp:pagination ?pages;
        dblp:yearOfPublication ?year;
        dblp:publishedAsPartOf <__proceedings_uri__> .
    FILTER (STRSTARTS(str(?doi), "https://doi.org/"))
}

# get the list of authors associated in papers of a given
# proceeding identifier <__proceedings_uri__>

PREFIX dblp: <https://dblp.org/rdf/schema#> 

SELECT ?paper ?title ?name ?ordinal ?orcid ?wikidata ?scholar {
    ?paper a dblp:Publication, dblp:Inproceedings;
        dblp:title ?title;
        dblp:hasSignature ?sign;
        dblp:publishedAsPartOf <__proceedings_uri__> .

    ?sign dblp:signatureDblpName ?name;
        dblp:signatureCreator ?dblp_person;
        dblp:signatureOrdinal ?ordinal .

    OPTIONAL { ?dblp_person dblp:orcid ?orcid }
    OPTIONAL { ?dblp_person dblp:wikidata ?wikidata }
    OPTIONAL { ?dblp_person dblp:webpage ?scholar . 
    FILTER (STRSTARTS(str(?scholar),''https://scholar.google.com/")) }
}
\end{lstlisting}
\label{lst:ex3}
\end{figure}

\section{OpenRefine Schema for Scholarly Article Type}\label{app:open_refine_view}

\begin{figure}[h]
  \centering
  \includegraphics[width=0.78\textwidth]{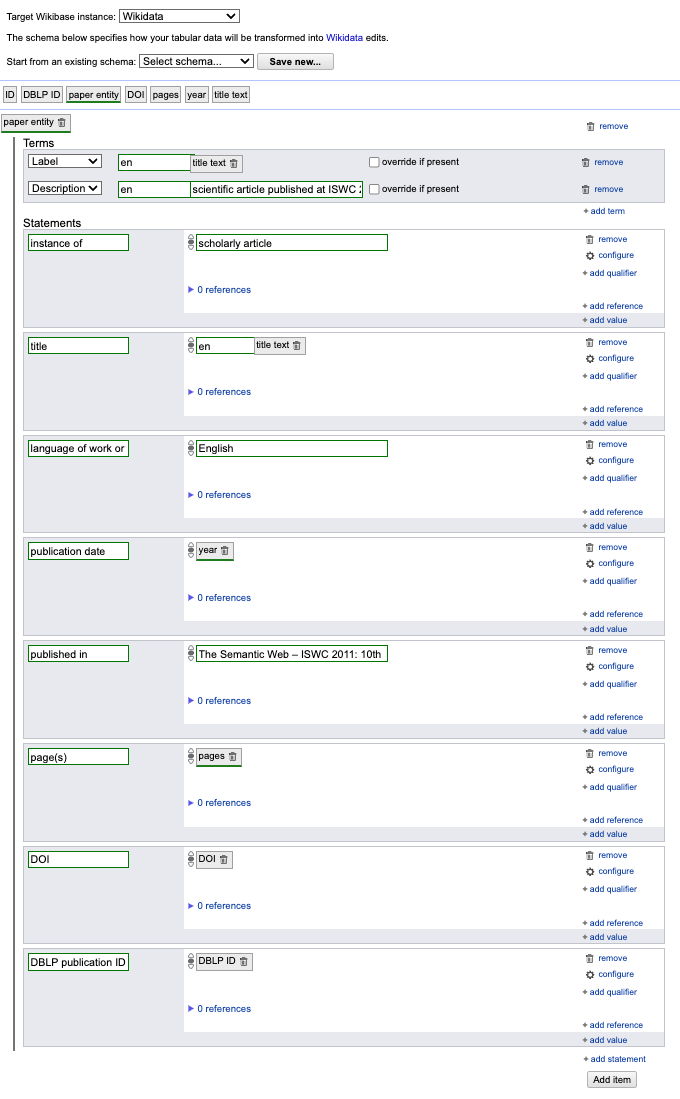}
  \caption{Screenshot Open Refine schema that maps the papers extracted from the SPARQL query to Wikidata for generating Quick Statements.}
  \label{fig:open-refine-screenshot}
\end{figure}

\end{document}